\documentclass{amsart}
\usepackage[centertags]{amsmath}
\usepackage{amsfonts}
\usepackage{amssymb}
\usepackage{amsthm}
\usepackage{amsmath}
\usepackage{amscd}
\usepackage{latexsym}
\usepackage{color}
\usepackage{graphics}

\newtheorem{thm}{Theorem}[section]
\newtheorem{cor}[thm]{Corollary}

\newtheorem{prop}[thm]{Proposition}
\theoremstyle{definition}
\theoremstyle{remark}
\numberwithin{equation}{section}

\DeclareMathOperator{\RE}{Re} \DeclareMathOperator{\IM}{Im}

\newcommand{\ep}{\varepsilon}

\newcommand{\vph}{\varphi}

\newcommand{\intl}{\int\limits}

\DeclareMathAccent{\acute}{\mathalpha}{operators}{"13}
\DeclareMathAccent{\grave}{\mathalpha}{operators}{"12}
\DeclareMathAccent{\ddot}{\mathalpha}{operators}{"7F}
\DeclareMathAccent{\tilde}{\mathalpha}{operators}{"7E}
\DeclareMathAccent{\bar}{\mathalpha}{operators}{"16}
\DeclareMathAccent{\breve}{\mathalpha}{operators}{"15}
\DeclareMathAccent{\check}{\mathalpha}{operators}{"14}
\DeclareMathAccent{\hat}{\mathalpha}{operators}{"5E}
\DeclareMathAccent{\vec}{\mathord}{letters}{"7E}
\DeclareMathAccent{\dot}{\mathalpha}{operators}{"5F}
\DeclareMathAccent{\widetilde}{\mathord}{largesymbols}{"65}
\DeclareMathAccent{\widehat}{\mathord}{largesymbols}{"62}
\begin{document}
\title
 {On some discrete model of the magnetic Laplacian}

\author{ Volodymyr Sushch }

\address{Department of Mathematics, Technical University of Koszalin, Sniadeckich 2,
 75-453 Koszalin, Poland;  Pidstrygach Institute for Applied Problems of Mechanics and
 Mathematics,  Lviv, Ukraine }

\email{sushch@lew.tu.koszalin.pl}



\date{Octovber 2004}
\maketitle
\begin{abstract}
We construct some intrinsically defined discrete model of the
magnetic Laplacian. The existence and uniqueness of solutions of
the Dirichlet problem for the difference Poisson type equation are
proved. We study in detail properties of the discrete model
including the limiting process in the two-dimensional Euclidean
case.

\end{abstract}

\section{Introduction}
Let $(M, g)$ be a Riemannian manifold with a Riemannian metric \
$(g_{ij})$ \ and \ $dimM=n$. Denote $\Lambda^p(M)$ the set of all
differentiable  complex-valued  $p$-forms on $M$ for each $p=0,1,
..., n$. Note that $\Lambda^0(M)$ is just $C^{\infty}(M)$. Denote
also $\Lambda^p_{(k)}(M)$ the set of all $k$-smooth (of the class
$C^k$)complex-valued  $p$-forms on $M$. We define a {\it magnetic
potential} as a real-valued 1-form $A\in \Lambda^1_{(1)}(M)$. So
in local coordinates $x_1, ..., x_n$ it can be written as
$$A=\sum_{j=1}^nA_jdx^j,$$ where $A_j=A_j(x)$ are real-valued
functions of the class $C^1$.

Let $\ast$ be the metric adjoint operator (Hodge star operator)
$\ast: \Lambda^p(M)\rightarrow \Lambda^{n-p}(M)$. Then the
invariant inner product of $p$-forms with compact support can be
defined in a standard way by the relation
\begin{equation} \label{1.1}(\vph, \
\psi)=\intl_M\vph\wedge\ast\overline{\psi},
\end{equation}
 where the bar
over $\psi$ means the complex conjugation and $\wedge$ is the
exterior multiplication operation. It is known that using the
inner product (\ref{1.1}) in spaces of smooth $p$-forms with
compact support we can define the completions of these spaces.  We
denote these Hilbert spaces by $L^2(M)$ for 0-forms (functions)
and by $L^2\Lambda^p(M)$ for $p$-forms, $p=1, ..., n$. Let us
define the operator $$d:L^2\Lambda^{p-1}(M)\rightarrow
L^2\Lambda^{p}(M)$$ as the closure in the $L^2$-norm generated by
the inner product (\ref{1.1}) of the corresponding operator
specified on smooth forms, i.e. as a strong extensions of the
differential operator $d:\Lambda^{p-1}(M)\rightarrow
\Lambda^{p}(M).$ We will need a deformed differential
 \begin{equation} \label{1.2}
 d_A:L^2(M)\rightarrow
L^2\Lambda^{1}(M), \qquad \vph\rightarrow d\vph+i\vph A,
 \end{equation}
 where
$i=\sqrt{-1}$ and $A$ is the magnetic potential.

The definition of the invariant inner product immediately induces
the formal adjoint operator to the differential operator $d_A$. So
we have the operator $$\delta_A:L^2\Lambda^{1}(M)\rightarrow
L^2(M)$$ giving by the relation $$(d_A\vph, \ \omega)=(\vph, \
\delta_A\omega), \qquad \vph\in L^2(M), \quad \omega\in
L^2\Lambda^{1}(M).$$ Here we assume that one of the forms have
compact support. Then we can define the {\it magnetic Laplacian}
as follows
 \begin{equation}
\label{1.3}-\Delta_A\equiv\delta_Ad_A:L^2(M)\rightarrow L^2(M).
\end{equation}
Let us identify the magnetic potential $A$ with the multiplication
operation
\begin{equation}
\label{1.4} A: L^2(M)\rightarrow L^2\Lambda^{1}(M), \qquad
\vph\rightarrow\vph A
\end{equation}(see e.g. \cite{Shubin}).
 Then
the formally adjoint operator $\delta_A$ can be written as follows
\begin{equation}
\label{1.5} \delta_A\omega=(\delta-iA^\ast)\omega,
\end{equation} where
$\delta, \ A^\ast$ are the formal adjoint operators to $d$ and $A$
respectively. Using (\ref{1.2}),(\ref{1.3}), we can rewrite the
magnetic Laplacian $\Delta_A$ as follows
\begin{align*}
-\Delta_A\vph &=(\delta-iA^\ast)(d\vph+iA\vph)\\ &=\delta
d\vph-iA^\ast d\vph+i\delta(A\vph)+A^\ast A\vph\\
&=-\Delta\vph-iA^\ast d\vph+i\delta(A\vph)+A^\ast A\vph.
\end{align*}
Operator (\ref{1.3}) is essentially self-adjoint (see for details
\cite[Th.~6.1]{Shubin}).

 The main purpose of this paper is to construct an
intrinsically defined discrete model of the magnetic Laplacian.
Speaking about this discrete model we do not mean just the
corresponding difference operator on a lattice or on graphs but we
mean a discrete analog of the Riemannian structure on some
combinatorial object. We consider discrete forms as certain
cochains. We construct discrete analogs of the exterior
multiplication operation, the Hodge star operator, of  inner
product (\ref{1.1}) and the operators (\ref{1.2}), (\ref{1.5}).

 Our approach bases on the formalism
proposed by Dezin \cite{Dezin}. For an account of other geometric
finite-difference approaches to Hodge theory of harmonic forms see
references \cite{Bak, Dod, Kom}. The discrete magnetic Laplacian
on graphs had been studied in \cite{BSBE}, \cite{DodM}.

In the spirit of \cite{Dezin}, \cite{S1}, \cite{S2} we study
self-adjointness of the discrete magnetic Laplacian and we proof
that the Dirichlet problem for the discrete Poisson type equation
has a unique solution.

In this paper we consider just the two dimensional Euclidean case.
Although similar constructions can be carried out in the
$n$-dimensional case, the two-dimensional discrete model makes it
possible to analyze in detail the combinatorial relations and the
limiting process.  One of the formal results is the construction
of a nonstandard approximation of the generalized solution of the
Poisson type equation for the magnetic Laplacian (\ref{1.3}) under
the minimal requirements of smoothness of the right hand side (is
belonged to $L^2$).

\section{Preliminaries on combinatorial structures}
We use the schema of discretization  due to Dezin \cite{Dezin}.
Let $\{x_k\}$, $\{e_k\}$, $k\in\mathbb{Z}$, be the sets of basis
elements of real linear spaces $C^0$, $C^1$. We will regard the
linear combinations $a=\sum a^kx_k$, \ $b=\sum b^kx_k$, \ $a^k,
b^k\in\mathbb{R}$, as zero-dimensional and one-dimensional chains,
respectively.

It is convenient to introduce shift operators $$\tau k=k+1, \qquad
\sigma k=k-1$$ in the set of indices. We define the
one-dimensional complex $C$ as the direct sum $C^0\oplus C^1$ of
the introduced spaces with the following boundary operator
$$\partial x_k=0, \qquad \partial e_k=x_{\tau k}-x_k, \qquad
k\in\mathbb{Z}.$$ The definition of $\partial$ is linearly
extended to arbitrary chains. We call the complex $C$ a
combinatorial model of the real line. The basis elements $x_k,
e_k$ can be interpreted as points and intervals connecting the
points (i.e. $e_k=(x_k, x_{\tau k})$) of real line.

We consider the tensor degree $C(n)=\otimes_1^n C$ of the
one-dimensional complex $C$ as a combinatorial model of
$\mathbb{R}^n$. The main object of our study will be a discrete
model of the magnetic Laplacian in the simplest two-dimensional
domain. Therefore we describe the combinatorial relations that are
encountered in the two-dimensional case.

The basis elements of the two-dimensional complex $C(2)$ can be
written as follows
\begin{align*}
x_k\otimes x_s=x_{k,s}, &\qquad e_k\otimes x_s=e_{k,s}^1,\\
e_k\otimes e_s=V_{k,s}, &\qquad x_k\otimes e_s=e_{k,s}^2.
\end{align*}
The boundary operator $\partial$ we define as
\begin{align}\notag
\partial x_{k,s}=0, &\qquad \partial e_{k,s}^1=x_{\tau k,s}-x_{k,s}, \qquad
\partial e_{k,s}^2=x_{k,\tau s}-x_{k,s}, \\
\label{2.1} &\partial V_{k,s}=e_{k,s}^1+e_{\tau k,s}^2-e_{k,\tau
s}^1-e_{k,s}^2.
\end{align}

Let us introduce an object dual to $C(2)$. Namely, the complex of
complex-valued functions over $C(2)$. The dual complex $K(2)$ we
can consider as the set of complex-valued cochains  and it has the
same structure as $C(2)$, i.e. $K(2)=K\otimes K$. In other words,
$K(2)$ is a linear complex space with basis elements $\{x^{k,s}, \
e^{k,s}_1, \ e^{k,s}_2, \ V^{k,s}\}$.

The pairing (chain-cochain) operation is defined by the rules:
\begin{equation}\label{2.2}
<x_{k,s}, \ x^{p,q}>=<e_{k,s}^1, \ e^{p,q}_1>=<e_{k,s}^2, \
e^{p,q}_2>=<V_{k,s}, \ V^{p,q}>=\delta_{k,s}^{p,q},
\end{equation}
where $\delta_{k,s}^{p,q}$ is  Kronecker symbol. We call elements
of the complex $K(2)$ forms. Then the 0-, 1-, 2-forms $\vph, \
\omega=(u, v), \ \eta$ can be written as
\begin{equation}\label{2.3}
\vph=\sum_{k,s}\vph_{k,s}x^{k,s}, \quad
\omega=\sum_{k,s}(u_{k,s}e^{k,s}_1+v_{k,s}e^{k,s}_2), \quad
\eta=\sum_{k,s}\eta_{k,s}V^{k,s},
\end{equation}
 where $\vph_{k,s}, \ u_{k,s}, \
v_{k,s},  \ \eta_{k,s}\in\mathbb{C}$ for any $k,s\in\mathbb{Z}$.
 The pairing (\ref{2.2}) is
linearly extended to forms (\ref{2.3}). The boundary operation
$\partial$ in $C(2)$ (\ref{2.1}) induces the dual operation $d^c$
in $K(2)$:
\begin{equation}\label{2.4}
<\partial a, \ \alpha>=<a, \ d^c\alpha>,
\end{equation}
where $a\in C(2), \ \alpha\in K(2)$. The coboundary  operator
$d^c$ is a discrete analog of the exterior differentiation
operator $d$. We will need the expression for $d^c$ over the basis
elements of $C(2)$:
\begin{align}\notag
 &<e^1_{k,s}, \ d^c\vph>=\vph_{\tau k,
 s}-\vph_{k,s}\equiv\Delta_k\vph_{k,s},\\
 \label{2.5}
 & <e^2_{k,s}, \ d^c\vph>=\vph_{k, \tau
 s}-\vph_{k,s}\equiv\Delta_s\vph_{k,s},\\ \notag
  &<V_{k,s}, \ d^c\omega>=v_{\tau k,
 s}-v_{k,s}-u_{k,\tau s}+u_{k,s}\equiv\Delta_k v_{k,s}-\Delta_s
 u_{k,s}.
 \end{align}

 We define the multiplication $\cup$ in  $K(2)$ by the rules:
\begin{align}\label{2.6}\notag
&x^{k,s}\cup x^{k,s}=x^{k,s}, \qquad e^{k,s}_2\cup e^{k,\tau
s}_1=-V^{k,s},\\  &x^{k,s}\cup e^{k,s}_1=e^{k,s}_1\cup x^{\tau
k,s}=e^{k,s}_1, \qquad x^{k,s}\cup e^{k,s}_2=e^{k,s}_2\cup
x^{k,\tau s}=e^{k,s}_2,
\\ \notag &x^{k,s}\cup V^{k,s}= V^{k,s}\cup x^{\tau k,\tau s}=e^{k,s}_1\cup
e^{\tau k,s}_2=V^{k,s},
 \end{align}
 supposing the product to be zero in all other cases. To
 forms (\ref{2.3}) the $\cup$-multi\-plication can be extended
 linearly. For arbitrary forms $\alpha, \beta\in K(2)$ we have
 (see \cite[p. 147]{Dezin}) the relation
 \begin{equation}\label{2.6a}
d^c(\alpha\cup\beta)=d^c\alpha\cup\beta+(-1)^r\alpha\cup d^c\beta,
\end{equation}
where $r$ is the dimension of the form $\alpha$. So the
$\cup$-multiplication is an analog of the exterior multiplication
$\wedge$ for differential forms.

Let $\ep^{k,s}$ be an arbitrary basis elements of $K(2)$. We
introduce the "star"  operator setting
\begin{equation}\label{2.7}
\ep^{k,s}\cup\ast\ep^{k,s}=V^{k,s}.
\end{equation}
Using (\ref{2.6}), we have
\begin{equation*}
\ast x^{k,s}=V^{k,s},  \quad \ast e^{k,s}_1=e^{\tau k,s}_2,  \quad
\ast e^{k,s}_2=-e^{k,\tau s}_1,  \quad \ast V^{k,s}=x^{\tau k,\tau
s}.
\end{equation*}
The operator $\ast$ is extended to arbitrary forms by linearity.

Let now
\begin{equation}\label{2.8}
V=\sum_{k,s}V_{k,s},  \qquad k=1, 2, ..., N, \quad s=1, 2, ..., M
\end{equation}
is some fixed "domain", namely, a set of 2-dimensional basis
elements of $C(2)$. Then the relation
\begin{equation}\label{2.9}
(\alpha, \ \beta)_V=<V, \ \alpha\cup\ast\overline{\beta}>
\end{equation}
gives a correct definition of the inner product for forms of the
same degree (cf. (\ref{1.1})). For forms of different degree the
product (\ref{2.9}) is equal to zero. Using
(\ref{2.6})--(\ref{2.8}), we obtain
\begin{equation}\label{2.10}
(\alpha, \ \beta)_V=\sum_{k,s}\alpha_{k,s}\overline{\beta_{k,s}},
\end{equation}
where $\alpha_{k,s},\ \beta_{k,s}$ are components of the forms
$\alpha,\ \beta\in K(2)$.

 We agree that in what follows, unless
the limits of summation are specified, the subscripts $k,s$ always
run over the set of values indicated in (\ref{2.8}).

Taking into account (\ref{2.6a}), (\ref{2.9}), we can written for
a $(p-1)$-form $\alpha\in K(2)$ and  $p$-form $\beta\in K(2)$ the
relation
\begin{equation}\label{2.11}
(d^c\alpha, \ \beta)_V=<\partial V, \
\alpha\cup\ast\overline{\beta}>+(\alpha, \ \delta^c\beta)_V,
\end{equation}
where
\begin{equation}\label{2.12}
\delta^c\beta=(-1)^p\ast^{-1}d^c\ast\beta.
\end{equation}
Here $\ast^{-1}$ is the operation inverse to $\ast$, i.e.
$\ast^{-1}\ast=1$.  If the form $\alpha\cup\ast\overline{\beta}$
vanishes on the boundary $\partial V$, then Equation (\ref{2.12})
defines the formally adjoint operator to $d^c$. Let $\omega=(u,v)$
be an 1-form (\ref{2.3}). Then we have
\begin{equation}\label{2.13}
\delta^c\omega=\sum_{k,s} (-\Delta_k u_{\sigma k,s}-\Delta_s
v_{k,\sigma s})x^{k,s}.
\end{equation}
We call the operator $\delta^c$ a discrete analog of the
codifferential $\delta$.

Therefore a discrete analog of the Laplace operator can be defined
as follows
\begin{equation*}\label{2.14}
-\Delta^c=\delta^c d^c+d^c\delta^c.
\end{equation*}
If $\vph$ is a 0-form, then $-\Delta^c\vph=\delta^c d^c\vph$ and
we obtain at the point $x_{k,s}$  the difference expression
\begin{equation*}\label{2.15}
<x_{k,s}, \ -\Delta^c\vph>=4\vph_{k,s}-\vph_{\tau
k,s}-\vph_{k,\tau s}-\vph_{\sigma k,s}-\vph_{k,\sigma s}.
\end{equation*}

It should be noted that the definition of the inner product
(\ref{2.10}) turns the linear space of forms over $V$ into
finite-dimensional Hilbert spaces $H^0, \ H^1, \ H^2$ with bases
$\{x^{k,s}\}$, \ $\{e^{k,s}_1, e^{k,s}_2\}$, \  $\{V^{k,s}\}$, \
$k=1, 2, ..., N, \ s=1, 2, ..., M$, respectively. Thus we can
regard the operators $d^c, \ \delta^c, \ \Delta^c$  over $V$ as
follow
\begin{equation*}
 d^c: H^p\rightarrow H^{p+1}, \qquad \delta^c: H^p\rightarrow H^{p-1}, \qquad \Delta^c: H^p\rightarrow H^p,
\end{equation*}
where $p=0,1,2$. It is convenient to suppose that $H^{-1}=H^3=0$.

\section{Discrete model of the magnetic Laplacian}
Let a real-valued 1-form
\begin{equation*}
 A=\sum_{k,s}(A_{k,s}^{1}e_{1}^{k,s}+A_{k,s}^{2}e_{2}^{k,s}),
\end{equation*}
$A_{k,s}^{1}, A_{k,s}^{2}\in\mathbb{R},$ be a discrete analog of
the  {\it magnetic potential}. We define the discrete analog of
the deformed differential (\ref{1.2}) as follows
\begin{equation}\label{3.1}
d^c_A: H^0\rightarrow H^1, \qquad \vph\rightarrow
d^c\vph+i\vph\cup A.
\end{equation}
Taking into account (\ref{2.5}), (\ref{2.6}), we have
\begin{equation}\label{3.2}
d^c_A\vph=\sum_{k,s}\big((\Delta_k\vph_{k,s}+i\vph_{k,s}A_{k,s}^{1})e_{1}^{k,s}+
(\Delta_s\vph_{k,s}+i\vph_{k,s}A_{k,s}^{2})e_{2}^{k,s}\big).
\end{equation}
As in the continual case (see (\ref{1.4})), we can  identify the
discrete magnetic potential $A$ with the multiplication operator
\begin{equation}\label{3.3}
A:H^0\rightarrow H^1,  \qquad \vph\rightarrow\vph\cup A.
\end{equation}
Then we have
\begin{equation*}
 A\vph=\sum_{k,s}(\vph_{k,s}A_{k,s}^{1}e_{1}^{k,s}+\vph_{k,s}A_{k,s}^{2}e_{2}^{k,s}).
\end{equation*}
\begin{prop} \label{prop1}The formally adjoint operator   \
$A^\ast:H^1\rightarrow H^0$ acts on an arbitrary 1-form
$\omega=(u, v)$ as follows
\begin{equation}\label{3.4}
A^{\ast}\omega=\sum\limits_{k,s}(A^{1}_{k,s}u_{k,s}+A^{2}_{k,s}v_{k,s})x^{k,s}.
\end{equation}
\end{prop}
\begin{proof}
Since the 1-form  $A\in H^1$ is real-valued by assumption, we have
\begin{align*}
(A\vph, \ \omega)_V&=(\vph\cup A, \ \omega)_V=<V, \ (\vph\cup
A)\cup\ast\overline{\omega}>\\
&=\sum_{k,s}\big((\vph_{k,s}A_{k,s}^1)\overline{u_{k,s}}+(\vph_{k,s}A_{k,s}^2)\overline{v_{k,s}}\big)\\&=
\sum_{k,s}\vph_{k,s}(\overline{A_{k,s}^1u_{k,s}+A_{k,s}^2v_{k,s}})=(\vph,
\ A^\ast\omega)_V.
\end{align*}
\end{proof}
Let us suppose that  components \ $\alpha_{k,s}$ \ of an arbitrary
$r$-form \ $\alpha\in H^r$, \  $r=0,1,2$, satisfy the following
"boundary conditions":
\begin{equation}\label{3.5}
\alpha_{0,s}=\alpha_{\tau N,s}=0, \qquad
\alpha_{k,0}=\alpha_{k,\tau M}=0
\end{equation}
 for all $k=1,2, ..., N, \quad
s=1,2, ..., M$.
\begin{prop}\label{prop2}Let
components of  $\vph\in H^0, \ \omega\in H^1$ satisfy Conditions
(\ref{3.5}). Then
 $$(d^c_A\vph, \ \omega)_V=(\vph, \
\delta^c_A\omega)_V,$$ where
\begin{equation}\label{3.6}
\delta^c_A\omega=\delta^c\omega-iA^\ast\omega.
\end{equation}
\end{prop}
\begin{proof} Note that  Conditions (\ref{3.5}) imply the relation
$<\partial V, \ \vph\cup\ast\overline{\omega}>=0$
\cite[p.~161]{Dezin}. Then from (\ref{2.11}) we have
\begin{equation*}
(d^c\vph, \ \omega)_V=(\vph, \ \delta^c\omega)_V.
\end{equation*}
Hence
\begin{align*}
 (d^c_A\vph, \ \omega)_V&=(d^c\vph+i\vph\cup A, \
\omega)_V=(d^c\vph, \ \omega)_V+i(\vph\cup A, \
 \omega)_V\\
&=(\vph, \ \delta^c\omega)_V+i(\vph, \ A^\ast\omega)_V=(\vph,
\
 (\delta^c-iA^\ast)\omega)_V.
 \end{align*}

\end{proof}
 Thus the operator $\delta^c_A: H^1\rightarrow H^0$ is the formally adjoint operator to the
 operator $d^c_A$.
 Using (\ref{2.13}), we can rewrite (\ref{3.6}) in a "pointwise"
 form:
\begin{equation*}
<x_{k,s}, \ \delta^c_A\omega>=-\Delta_ku_{\sigma
k,s}-\Delta_sv_{k,\sigma s}-i(A_{k,s}^1u_{k,s}+A_{k,s}^2v_{k,s}).
\end{equation*}
We have
\begin{align*}
\vph\cup\delta^c\omega&=\sum_{k,s}\vph_{k,s}(-\Delta_ku_{\sigma
k,s}-\Delta_sv_{k,\sigma s})x^{k,s}\\&=\sum_{k,s}(\vph_{\sigma
k,s}u_{\sigma
k,s}-\vph_{k,s}u_{k,s}-\vph_{k,s}v_{k,s}+\vph_{k,\sigma
s}v_{k,\sigma s})x^{k,s}\\&+\sum_{k,s}(\vph_{k,s}u_{\sigma
k,s}-\vph_{\sigma k,s}u_{\sigma k,s}+\vph_{k,s}v_{k,\sigma
s}-\vph_{k,\sigma s}v_{k,\sigma s})x^{k,s}\\
&=\delta^c(\vph\cup\omega)+\sum_{k,s}\big((\Delta_k\vph_{\sigma
k,s})u_{\sigma k,s}+(\Delta_s\vph_{k,\sigma s})v_{k,\sigma
s}\big)x^{k,s}.
\end{align*}
It follows that
$$\delta^c(\vph\cup\omega)=\vph\cup\delta^c\omega-\sum_{k,s}\big((\Delta_k\vph_{\sigma
k,s})u_{\sigma k,s}+(\Delta_s\vph_{k,\sigma s})v_{k,\sigma
s}\big)x^{k,s}.$$ From this we immediately obtain the following
discrete Leibniz rule for $\delta^c_A$:
\begin{align*}
\delta^c_A(\vph\cup\omega)&=(\delta^c-iA^\ast)(\vph\cup\omega)=\delta^c(\vph\cup\omega)-iA^\ast(\vph\cup\omega)\\
&=\vph\cup\delta^c\omega-\vph\cup iA^\ast
\omega-\sum_{k,s}\big((\Delta_k\vph_{\sigma k,s})u_{\sigma
k,s}+(\Delta_s\vph_{k,\sigma s})v_{k,\sigma s}\big)x^{k,s}\\
&=\vph\cup\delta^c_A\omega-\sum_{k,s}\big((\Delta_k\vph_{\sigma
k,s})u_{\sigma k,s}+(\Delta_s\vph_{k,\sigma s})v_{k,\sigma
s}\big)x^{k,s}
\end{align*}
(cf. \cite[Sect.~2]{Shubin}, where the corresponding Leibniz rule
is given in the continual case).

Let us define the discrete magnetic Laplacian as
\begin{equation*}
-\Delta^c_A=\delta^c_A d^c_A: H^0\rightarrow H^0.
\end{equation*}
Note that we assume that Conditions (\ref{3.5}) are satisfied for
any form $\vph\in H^0$. This gives us the necessary extension of
$\vph$ beyond $H^0$ to consider the operator $-\Delta^c_A$ as
above.

Using (\ref{3.1}), (\ref{3.6}), we have
\begin{align}\label{3.7}\notag
-\Delta^c_A\vph&=\delta^c_A(d^c\vph+i\vph\cup A)\\\notag
&=(\delta^c-iA^\ast)d^c\vph+(\delta^c-iA^\ast)(i\vph\cup
A)\\\notag &=-\Delta^c\vph-iA^\ast d^c\vph+i\delta^c(\vph\cup A) +
A^\ast(\vph\cup A)\\ &=-\Delta^c\vph-iA^\ast
d^c\vph+i\delta^cA\vph + A^\ast A\vph.
\end{align}
\begin{prop} The operator  $-\Delta^c_A$ is self-adjoint, i.~e.
$$(\delta^c_A d^c_A\vph, \ \psi)_V=(\vph, \ \delta^c_A
d^c_A\psi)_V.$$
\end{prop}
\begin{proof}
It is known (see \cite[p.~163]{Dezin}) that under Conditions
(\ref{3.5}) the discrete Laplacian $-\Delta^c=\delta^c
d^c:H^0\rightarrow H^0$ is self-adjoint. Using Propositions
\ref{prop1}, \ref{prop2}, we obtain
\begin{align*}(\delta^c_A d^c_A\vph, \ \psi)_V &=(\delta^c
d^c\vph, \ \psi)_V-(iA^\ast d^c\vph, \ \psi)_V+(i\delta^cA\vph, \
\psi)_V+(A^\ast A\vph, \ \psi)_V \\&=(\vph, \ \delta^c d^c
\psi)_V+( d^c\vph, \ iA\psi)_V-(A\vph, \ id^c\psi)_V+(A\vph, \
A\psi)_V \\ &=(\vph, \ -\Delta^c\psi)_V+(\vph, \
i\delta^cA\psi)_V-(\vph, \ iA^\ast d^c\psi)_V+(\vph, \ A^\ast
A\psi)_V \\&=(\vph, \ (-\Delta^c+i\delta^cA-iA^\ast d^c+A^\ast
A)\psi)_V=(\vph, \ -\Delta^c_A \psi)_V.
 \end{align*}
\end{proof}
Using (\ref{3.1}), we can write
\begin{align}\label{3.80}\notag
 (d^c_A\vph, \ d^c_A\psi)_V&=(d^c\vph,
d^c\psi)_V+(d^c\vph, \ iA\psi)_V\\&+(iA\vph, \ d^c\psi)_V+(iA\vph,
\ iA\psi)_V.
\end{align}
Taking into account (\ref{2.5}) and (\ref{2.10}), we have
\begin{align*}(d^c\vph, \
A\psi)_V&=\sum_{k,s}\big(\Delta_k\vph_{k,s}(\overline{\psi_{k,s}}A^1_{k,s})+
\Delta_s\vph_{k,s}(\overline{\psi_{k,s}}A^2_{k,s})\big)\\
&=\sum_{k,s}\vph_{k,s}\big(-\Delta_k(\overline{\psi_{\sigma
k,s}}A^1_{\sigma k,s})-\Delta_s(\overline{\psi_{k,\sigma
s}}A^2_{k,\sigma s})\big)\\ &+\sum_s\big(\vph_{\tau
N,s}(\overline{\psi_{N,s}}A^1_{N,s})-\vph_{1,s}(\overline{\psi_{0,s}}A^1_{0,s})\big)
\\&+\sum_k\big(\vph_{k,\tau
M}(\overline{\psi_{k,M}}A^2_{k,M})-\vph_{k,1}(\overline{\psi_{k,0}}A^2_{k,0})\big)
\\&=\sum_s\big(\vph_{\tau
N,s}(\overline{\psi_{N,s}}A^1_{N,s})-\vph_{1,s}(\overline{\psi_{0,s}}A^1_{0,s})\big)\\&+\sum_k\big(\vph_{k,\tau
M}(\overline{\psi_{k,M}}A^2_{k,M})-\vph_{k,1}(\overline{\psi_{k,0}}A^2_{k,0})\big)
+(\vph, \ \delta^c A\psi)_V.
\end{align*}
It follows that
\begin{align*}
(d^c_A\vph, \ d^c_A\psi)_V&=\sum_s\big[\vph_{\tau
N,s}\big(\overline{\psi_{\tau
N,s}}-\overline{\psi_{N,s}}-i\overline{\psi_{N,s}}A^1_{N,s}\big)\big]\\&-\sum_s\big[
\vph_{1,s}\big(\overline{\psi_{1,s}}-\overline{\psi_{0,s}}+i\overline{\psi_{0,s}}A^1_{0,s}\big)\big]\\
&+\sum_k\big[\vph_{k,\tau M}\big(\overline{\psi_{k,\tau
M}}-\overline{\psi_{k,M}}-i\overline{\psi_{k,M}}A^2_{k,M}\big)\big]\\&-\sum_k\big[\vph_{k,1}\big(\overline{\psi_{k,1}}-\overline{\psi_{k,0}}+
i\overline{\psi_{k,0}}A^2_{k,0}\big)\big]\\&+(\vph,
-\Delta^c\psi)_V+(\vph,  i\delta^cA\psi)_V-(\vph,  iA^\ast
(d^c\psi))_V+(\vph,  A^\ast A\psi)_V.
\end{align*}
Consequently, if the forms $\vph, \psi\in H^0$ satisfy Conditions
(\ref{3.5}), then we obtain
\begin{equation}\label{3.8}(d^c_A\vph, \
d^c_A\psi)_V=-\sum_s\vph_{1,s}\overline{\psi_{1,s}}-\sum_k\vph_{k,1}\overline{\psi_{k,1}}+(\vph,
\ \delta^c_A d^c_A\psi)_V.
\end{equation}
\begin{thm} \label{thm1}For any form  $f\in H^0$ a solution of the equation
\begin{equation}\label{3.9}-\Delta^c_A \vph=f
\end{equation}
 exists and is unique.
\end{thm}
\begin{proof}
By virtue of the self-adjointness of the operator $-\Delta^c_A$ it
is enough to prove the uniqueness of the solution. Assume that
$\vph=\psi$ in Equation (\ref{3.8}). Then we can write
\begin{equation}\label{3.10} (d^c_A\vph, \
d^c_A\vph)_V+\sum_s|\vph_{1,s}|^2+\sum_k|\vph_{k,1}|^2=(\vph,
-\Delta^c_A \vph)_V.
\end{equation}
Using (\ref{2.5}), (\ref{2.10}), we get
\begin{equation*}
(d^c\vph, \
d^c\vph)_V=\sum_{k,s}\big(|\Delta_k\vph_{k,s}|^2+|\Delta_s\vph_{k,s}|^2\big),
\end{equation*}
\begin{align*}
(d^c\vph, \ iA\vph)_V+(iA\vph, \
d^c\vph)_V&=i\sum_{k,s}\big[A_{k,s}^1\big(\vph_{k,s}(\overline{\Delta_k\vph_{k,s}})-
(\Delta_k\vph_{k,s})\overline{\vph_{k,s}}\big)\big]\\
&+i\sum_{k,s}\big[A_{k,s}^2\big(\vph_{k,s}(\overline{\Delta_s\vph_{k,s}})-
(\Delta_s\vph_{k,s})\overline{\vph_{k,s}}\big)\big],
\end{align*}
\begin{equation*}
(iA\vph, \
iA\vph)_V=\sum_{k,s}\big((A_{k,s}^1)^2|\vph_{k,s}|^2+(A_{k,s}^2)^2|\vph_{k,s}|^2\big).
\end{equation*}
It is easy to check that
\begin{equation*}
\vph_{k,s}(\overline{\Delta_k\vph_{k,s}})-
(\Delta_k\vph_{k,s})\overline{\vph_{k,s}}=2i\big(\IM(\vph_{k,s})\RE(\Delta_k\vph_{k,s})-
\RE(\vph_{k,s})\IM(\Delta_k\vph_{k,s})\big).
\end{equation*}
Substituting the last relations into (\ref{3.80}) we obtain
\begin{align*}
 (d^c_A\vph, \ d^c_A\vph)_V=\sum_{k,s}\Big[|\Delta_k\vph_{k,s}|^2+|\Delta_s\vph_{k,s}|^2
 +(A_{k,s}^1)^2|\vph_{k,s}|^2+(A_{k,s}^2)^2|\vph_{k,s}|^2\\
+2A_{k,s}^1\RE(\vph_{k,s})\IM(\Delta_k\vph_{k,s})-2A_{k,s}^1\IM(\vph_{k,s})\RE(\Delta_k\vph_{k,s})\\
+2A_{k,s}^2\RE(\vph_{k,s})\IM(\Delta_s\vph_{k,s})-2A_{k,s}^2\IM(\vph_{k,s})\RE(\Delta_s\vph_{k,s})\Big]
 \end{align*}
\begin{align*}
 =\sum_{k,s}\Big[\Big(\RE(\Delta_k\vph_{k,s})-A_{k,s}^1\IM(\vph_{k,s})\Big)^2+
\Big(\IM(\Delta_k\vph_{k,s})+A_{k,s}^1\RE(\vph_{k,s})\Big)^2\\
\qquad\quad+\Big(\RE(\Delta_s\vph_{k,s})-A_{k,s}^2\IM(\vph_{k,s})\Big)^2+
\Big(\IM(\Delta_s\vph_{k,s})+A_{k,s}^2\RE(\vph_{k,s})\Big)^2\Big].
 \end{align*}

 Now let we take $f=0$ in Equation (\ref{3.9}). Then comparing the
 last equation and (\ref{3.10}), we obtain
\begin{align*}
 &\sum_{k,s}\Big[\Big(\RE(\Delta_k\vph_{k,s})-A_{k,s}^1\IM(\vph_{k,s})\Big)^2+
\Big(\IM(\Delta_k\vph_{k,s})+A_{k,s}^1\RE(\vph_{k,s})\Big)^2\\
&\quad+\Big(\RE(\Delta_s\vph_{k,s})-A_{k,s}^2\IM(\vph_{k,s})\Big)^2+
\Big(\IM(\Delta_s\vph_{k,s})+A_{k,s}^2\RE(\vph_{k,s})\Big)^2\Big]\\
+&\sum_s|\vph_{1,s}|^2+\sum_k|\vph_{k,1}|^2=0.
 \end{align*}
 It follows that $\vph_{k,s}=0$ for any
 $k,s$. Hence $\vph\equiv0$.
 \end{proof}

This immediately implies the following statement:
\begin{cor} The operator $-\Delta_A^c$ is positiv.
\end{cor}

\section {Approximation and limiting process}
In this section we consider the relationship between the
combinatorial objects that we have described above and the
corresponding continual objects. We will construct some
nonstandard approximation of the generalized solution of the
Poisson type equation
\begin{equation}\label{4.1}
-\Delta_A\vph=f,
\end{equation}
 where $f\in L^2(\Omega)$.  We will realize the scheme similar
 to that given in \cite[Ch.3, Sec.3]{Dezin}.

 Let the domain \ $\Omega\in \mathbb{R}^2$ \ be a rectangle with
 vertices \ $(a_1, \ b_1), \  (a_2, \ b_1)$, \  $(a_1, \ b_2), \  (a_2, \ b_2),$
where $0\leq a_1<a_2, \ 0\leq b_1<b_2$. We introduce a scale $h$
setting  $h=N^{-1}(a_2-a_1)=M^{-1}(b_2-b_1).$ Divide $\Omega$ by
the following straight lines $$x=a_1+kh, \quad y=b_1+sh, \quad
k=0,1, ..., N, \ s=0,1, ..., M.$$ We denote by $x_{k,s}$ the point
of intersection of these lines. We denote by $V_{k,s}$ an open
square bounded by the lines:  $x=a_1+kh, \ y=b_1+sh, \ x=a_1+\tau
kh, \ y=b_1+\tau sh.$ Let $e^1_{k,s}$ and $e^2_{k,s}$ be the
horizontal and vertical sides of $V_{k,s}$, i.~e.
$e^1_{k,s}=(x_{k,s}, x_{\tau k,s})$, \ $e^2_{k,s}=(x_{k,s},
x_{k,\tau s}).$ In this way we identify the rectangle $\Omega$
with the combinatorial domain $V$ (\ref{2.8}).

Let us now compare every discrete form $\vph\in H^0$ with the step
function assuming that $$\vph^h(x,y)=\vph_{k,s}, \qquad \mbox{for}
 \ x,y\in V_{k,s}.$$
 In the case of the 1-form \ $\omega=(u,v)\in H^1$ we have the pair of step
 functions $u^h(x,y)=u_{k,s}, \  v^h(x,y)=v_{k,s}$ and we can write
  $\omega^h(,x,y)=u^h(x,y)dx+v^h(x,y)dy$.
  Recall that $\vph_{k,s},u_{k,s}, v_{k,s}\in \mathbb{C}$
for any $k, s$. Then $\vph^h, \omega^h$ are complex-valued.

 It is easy to check that
 \begin{equation}\label{4.2}
\|\vph^h\|_{L^2(\Omega)}=h\|\vph\|_{H^0}, \qquad
\|\omega^h\|_{L^2\Lambda^1(\Omega)}=h\|\omega\|_{H^1}.
\end{equation}

Define difference operators acting on the step functions as
follows
\begin{align*}
\Delta_x^h\vph^h(x,y)=h^{-1}\big(\vph^h(x+h,y)-\vph^h(x,y)\big),\\
\Delta_y^h\vph^h(x,y)=h^{-1}\big(\vph^h(x,y+h)-\vph^h(x,y)\big).
\end{align*}
Replacing the partial derivatives $\frac{\partial}{\partial x}$, \
$\frac{\partial}{\partial y}$ appearing in $d, \ \delta$ by the
difference operators $\Delta_x^h, \ \Delta_y^h$, we can introduce
the difference operators $d^h, \ \delta^h$. The difference
equation
 \begin{equation}\label{4.3}
 d^h\vph^h=\omega^h
 \end{equation}
  is equivalent to the following
family of equations $$\Delta_k\vph_{k,s}=hu_{k,s}, \qquad
\Delta_s\vph_{k,s}=hv_{k,s},$$ where $k=0,1, ..., N, \ s=0,1, ...,
M.$ Hence Equation (\ref{4.3}) can be rewritten as the following
discrete equation $$d^c\vph=h\omega,$$ where $\vph, \ \omega$ are
discrete forms (see (\ref{2.3})) with the components $\vph_{k,s}$
and $u_{k,s}, v_{k,s}$, respectively. Similarly, we associate the
difference equation $\delta^h\omega^h=\vph^h$ and the discrete
equation $\delta^c\omega=h\vph$.

We can also introduce the following difference operator
\begin{equation}\label{4.4}
 d^h_{A^h}=d^h+iA^h,
  \end{equation}
  where $A^h=A^{1h}dx+A^{2h}dy$ and  $A^{1h}, A^{2h}$ are
  real-valued step functions defined as above. On the other hand,
   we can consider the step 1-form $A^h$ as the multiplication
   operator $A^h: L^2(\Omega)\rightarrow L^2\Lambda^1(\Omega)$
    acting as follows $$A^h\vph^h=\vph^hA^{1h}dx+\vph^hA^{2h}dy.$$
Then the formally adjoint operator to $A^h$ (cf. (\ref{3.4})) acts
on a step 1-form $\omega^h=(u^h, v^h)$ as
$$(A^h)^\ast\omega^h=u^hA^{1h}+v^hA^{2h}.$$ Thus we define the
difference magnetic Laplacian (cf. (\ref{3.7})) by the formula
\begin{equation}\label{4.5}
-\Delta^h_{A^h}\equiv\delta^h_{A^h}d^h_{A^h}=\delta^hd^h-i(A^h)^\ast
d^h+i\delta^hA^h + (A^h)^\ast A^h.
\end{equation}

 Now we consider the {\it discretization} procedure (see for
details \cite[p.~170]{Dezin}). Let $f(x,y)$ be a complex-valued
function defined over $\Omega$ (or over $V=\sum V_{k,s}$) and let
$f\in L^2(\Omega)$. Associate $f$ with the step function $f^h$
setting
\begin{equation}\label{4.6}
f^h(x,y)=h^{-2}\intl_{V_{k,s}}f(\xi,\eta)d\xi d\eta, \qquad
\mbox{for} \quad x,y\in V_{k,s}.
\end{equation}
Moreover, the value of $f^h$ can be assigned to the point
$x_{k,s}$. As above, we can write $f^h(x,y)=f_{k,s}$ for $x,y\in
V_{k,s}$. Thus we obtain the discrete 0-form $\widehat{f}=\sum
f_{k,s}x^{k,s}$ which is associating with  $f\in L^2(\Omega)$.
Similarly, if $f$ is an 1-form,  $f\in L^2\Lambda^1(\Omega)$, then
we associate each component of $f$ with the step function
(\ref{4.6}) and we assign the value of it to one of the intervals
$e_{k,s}^1$ or $e_{k,s}^2$.

Let us introduce the norm
\begin{equation*}
\parallel \vph^h\parallel_{W(\Omega)}^2=\intl_{\Omega}\big(|\Delta_x^h\vph^h|^2
 +|\Delta_y^h\vph^h|^2\big)dxdy.
\end{equation*}
It is not difficult to verify that
\begin{equation}\label{4.7}
\parallel \vph^h\parallel_{W(\Omega)}^2=\sum_{k=0}^N\sum_{s=0}^M\big(|\Delta_k\vph_{k,s}|^2
 +|\Delta_s\vph_{k,s}|^2\big).
\end{equation}
Hence we can write $$\parallel
\vph^h\parallel_{W(\Omega)}=\parallel \vph\parallel_{W(V)}.$$
\begin{thm}\label{thm2} Let the step function $f^h$ be the discretization of
$f\in L^2(\Omega)$. Then the following Dirichlet problem
\begin{equation}\label{4.8}
-\Delta^h_{A^h}\vph^h=f^h,
\end{equation}
\begin{equation}\label{4.9}
\vph^h|_{\partial\Omega}=0
\end{equation}
has a unique  solution and the inequality
\begin{equation}\label{4.10}
\parallel\vph^h\parallel_{W(\Omega)}<c_1\|\RE f\|_{L^2(\Omega)}+c_2\|\IM f\|_{L^2(\Omega)}
\end{equation}
is valid for the solution $\vph^h$.
\end{thm}
\begin{proof}
Using (\ref{4.5}), Equation (\ref{4.8}) can  be rewritten as
\begin{equation}\label{4.11}
\delta^hd^h\vph^h-i(A^h)^\ast d^h\vph^h+i\delta^hA^h\vph^h +
(A^h)^\ast A^h\vph^h=f^h.
\end{equation}
By definition the step function $\vph^h$ and the step form $A^h$
on $\Omega$ associated  with the discrete forms $\vph$ and $A$ on
$V$. As above, if we replace the difference operator $d^h,
\delta^h$ by the discrete operator $d^c, \delta^c$, then Equation
(\ref{4.11}) transforms into the following equation
\begin{equation}\label{4.12}
\delta^cd^c\vph-ihA^\ast d^c\vph+ih\delta^cA\vph +h^2A^\ast
A\vph=h^2\widehat{f},
\end{equation}
where $\widehat{f}$ is a 0-form defined by the step function $f^h$
(see (\ref{4.6})). Note that, if the step function $\vph^h$
satisfies Condition (\ref{4.9}), then the corresponding discrete
form satisfies Condition (\ref{3.5}). Thus the unique solvability
of (\ref{4.8}), (\ref{4.9}) immediately follows from Theorem
\ref{thm1}.

Let now represent Equation (\ref{4.8}) as follows
\begin{equation*}
-\Delta^h_{A^h}\RE\vph^h=\RE f^h, \qquad
-\Delta^h_{A^h}\IM\vph^h=\IM f^h.
\end{equation*}
In a similar way we can split Equation (\ref{4.12}).

 Since
\begin{equation*}
 (d^c\alpha, \ iA\alpha)_V+(iA\alpha, \ d^c\alpha)_V=0
 \end{equation*} for any
real-valued discrete form $\alpha\in H^0$
 it follows that from (\ref{3.7}), (\ref{3.10}) we obtain
\begin{equation*}
 \|d^c\RE\vph\|_{H^1}^2+\sum_s
 (\RE\vph_{1,s})^2+\sum_k(\RE\vph_{k,1})^2+h^2\|A\RE\vph\|_{H^1}^2=h^2(\RE\vph,\
\RE\widehat{f})_V.
\end{equation*}
It immediately follows that
\begin{equation*}
\|\RE\vph\|_{W(V)}^2<h^2(\RE\vph,\
\RE\widehat{f})_V=h^2\sum_{k,s}\RE\vph_{k,s}\RE f_{k,s}.
\end{equation*}
Hence, using (\ref{4.2}), we have
\begin{equation}\label{4.13}
\|\RE\vph^h\|_{W(\Omega)}^2<\|\RE\vph^h\|_{L^2(\Omega)}\|\RE
f^h\|_{L^2(\Omega)}.
\end{equation}
It is easy to check the following estimates
\begin{equation*}
\|\RE\vph^h\|_{L^2(\Omega)}\leq\|\RE\vph^h\|_{W(\Omega)}, \qquad
\|\RE f^h\|_{L^2(\Omega)}\leq c_1\|\RE f\|_{L^2(\Omega)}
\end{equation*}
(see for details \cite[Ch.3, Theorem~5]{Dezin}). Combining the
last with (\ref{4.13}), we obtain
\begin{equation}\label{4.14}
\|\RE\vph^h\|_{W(\Omega)}< c_1\|\RE f\|_{L^2(\Omega)}.
\end{equation}

Similarly we obtain the estimate
\begin{equation}\label{4.15}
\|\IM\vph^h\|_{W(\Omega)}<c_2\|\IM f\|_{L^2(\Omega)}.
\end{equation}

Finally since
\begin{equation*}
\|\vph^h\|_{W(\Omega)}\leq
\|\RE\vph^h\|_{W(\Omega)}+\|\IM\vph^h\|_{W(\Omega)}
\end{equation*}
Estimates (\ref{4.14}), (\ref{4.15}) imply (\ref{4.10}).
\end{proof}

Let us now consider the limiting process. As in
\cite[Ch.3]{Dezin}, by the step function $\vph^h$ we construct the
smooth (i.e. of the class $C^1$) function $J^h\vph^h$. It is
convenient to take $J^h\vph^h$ in the form
\begin{equation*}
J^h\vph^h(x,y)=h^{-2}\intl_x^{x+h}\intl_y^{y+h}\vph^h(\xi,\eta)d\xi
d\eta.
\end{equation*}
This is the well known Steklov function with the averaging radius
equal to the parameter h (the scale of the net). We can also
define the 1-form $J^hA^h\in \Lambda^1_{(1)}(\Omega)$ as
\begin{equation*}
J^hA^h(x,y)=J^hA^{1h}(x,y)dx+J^hA^{2h}(x,y)dy.
\end{equation*}

Denote by $\dot{W}^1(\Omega)$ the Sobolev space of complex-valued
function which satisfy the homogeneous Dirichlet condition.

Let us consider some sequence $\{h_n\}$ such that $h_n\rightarrow
0$ as $n\rightarrow+\infty$. For convenient  further we will write
$h$ instead $h_n$.

Let $A\in \Lambda^1_{(1)}(\Omega)$ in the operator $-\Delta_A$. We
have the statement.
\begin{thm} Let a step function $\vph^h$ be the solution of the
Dirichlet problem  (\ref{4.8}), (\ref{4.9}) for the given element
$f\in L^2(\Omega)$. Then the sequence $\{\vph^h\}$ strongly
converges in $L^2(\Omega)$ to the element
$\vph\in\dot{W}^1(\Omega)$ as $h\rightarrow 0$, where $\vph$ is
the generalized solution of the corresponding Dirichlet problem
for Equation (\ref{4.1}). At the same time the sequence
$\{J^h\vph^h\}$ converges to $\vph$ in the metric
$\dot{W}^1(\Omega)$.
\end{thm}
\begin{proof}
Based on Theorem \ref{thm2}, the proof is similar to that of
Theorem~5 \cite[ch.3]{Dezin}.
\end{proof}

\end{document}